\documentstyle[aps,prb,epsfig,multicol,color]{revtex}
\begin{document}
\draft
\title{Nuclear spin-lattice relaxation rate in the D+iD superconducting state: implications for CoO
superconductor}

\author{Yunkyu Bang$^{*}$, M. J. Graf, and A.V. Balatsky}
\address {Los Alamos National Laboratory, Los Alamos,
New Mexico 87545 }
\date{\today}
\maketitle

\begin{abstract}
We calculated the nuclear spin-lattice relaxation rate $1/T_1$ for
the D+iD superconducting state with impurities. We found that
small amount of unitary impurities quickly produces the residual
density of states inside the gap. As a result, the T-linear
behavior in 1/T$_1$ is observed at low temperatures. Our results
show that the D+iD pairing symmetry of the  superconducting state
of  Na$_{0.35}$CoO$_{2} \cdot y$H$_2$ O is  compatible with recent
$^{59}$Co 1/T$_1$ experiments of several groups.
\end{abstract}

\begin{multicols}{2}

\section{Introduction}

The recent discovery of Na$_{0.35}$CoO$_{2} \cdot y$H$_2$ O
superconductor \cite{discovery}  has renewed interest in the study
of exotic superconductivity (SC). Motivated by the triangular
lattice structure of spin-$\frac{1}{2}$ Co atoms of the reference
CoO$_2$ layer, several authors \cite{spin_liquid} have quickly
proposed that it might be a realization of the long sought
spin-liquid based superconductivity, originally proposed for
high-T$_c$ superconductors by Anderson et al.\cite{Anderson}.
Indeed several theoretical calculations \cite{D+iD} indicate that
the frustration of the antiferromagnetism (AFM)  on the triangular
lattice of Co favors the D+iD superconducting symmetry.

For a new superconducting material, the first step toward laying
down the theory of the  superconducting state is  the experimental
identification of the SC gap symmetry. Nuclear spin-lattice
relation rate 1/T$_1$ and Knight shift measurements have been the
successful tools for this purpose and  several measurements of
1/T$_1$ and Knight shift of this compound have already been
reported. At present, however, these reports appear inconsistent
with each other. Regarding the coherence peak of Co$^{59}$
1/T$_1$, for example, Kobayashi et al. \cite{Kobayashi} and Waki
et al. \cite{Waki} reported the existence of a coherence peak; the
maximum value of 1/T$_1$T is about 1.1 to 1.5 times of the normal
state value while an ideal s-wave case has almost twice of the
normal state value\cite{Tinkham}.  However, Fujimoto et al.
\cite{Fujimoto} recently reported no coherence peak. Furthermore,
they reported more precise  $^{59}$ Co 1/T$_1$ below T$_c$ and
found a substantial region of linear temperature dependence in
1/T$_1$ at low temperatures and T$^3$ behavior near  and below
T$_c$. To fit their data a residual density of states (DOS) of
0.65N(0) was needed assuming an impure D-wave SC state. From this
calculation, these authors concluded that the non S-wave gap with
the lines of nodes is most consistent with their 1/T$_1$ data and
the compatibility of the D+iD symmetry gap having a full gap
remains as a question.

In this note, we calculated 1/T$_1$ for D+iD gap with impurities
using self-consistent T-matrix approximation and quantitatively
compared our results with experiments reported up to date.
Since the impurity effects on various symmetry gaps has been
studied by many authors \cite{Hirschfeld,Norman} and the generic
behaviors are well known, we summarize them. Within T-matrix
approximation, the impurity scattering renormalizes the
self-energies both of normal and anomalous components. In Nambu
notation, they correspond to $\Sigma_0, \Sigma_2$, and $\Sigma_3$,
respectively; $\Sigma_0$ and $\Sigma_3$ are the two separate
components of the normal self-energy according to  particle-hole
transformation and $\Sigma_2$ is the anomalous self-energy
component.
Among them, the renormalization of $\Sigma_3$ usually vanishes for
particle-hole symmetric band, which is a good approximation for
most  cases. For an isotropic s-wave gap, both $\Sigma_0$ and
$\Sigma_2$ are renormalized and  their effects largely cancel each
other  in the renormalized gap equation; as a result the impurity
effect is null   for low temperature SC properties such as the
changes of the order parameter (OP) magnitude ($\Delta_0$) and
T$_c$, and the residual DOS etc. For a sign changing OP such as
the D-wave gap, the renormalization of $\Sigma_2$ vanishes, too,
due to symmetry, and the impurity effects appear only in the
renormalization of $\Sigma_0$ the normal self-energy. The case of
D+iD gap is interesting in that it has a full gap all over the FS
and should display similar thermodynamic behaviors as in an
isotropic s-wave gap. However, regarding the impurity effects, it
is expected to show a mixed behavior of a D-wave gap and a s-wave
gap because the sign changing OP leads to a vanishing $\Sigma_2$
renormalization despite a full gap.

We find that for unitary impurities the  residual DOS is created
in the D+iD gap as easily as in the D-wave gap. This happens
because the resonant scattering process of unitary impurity is
more efficient at lower DOS. As a result, D+iD gap  displays a
similar T-linear behavior in 1/T$_1$ at low temperatures with the
same amount of unitary impurities as in the simple D-wave case.
Therefore, the observation of the T-linear behavior in 1/T$_1$ at
low temperatures\cite{Fujimoto} itself cannot rule out the D+iD
symmetry gap on CoO$_2$ superconductor. Moreover, in contrast to
the D-wave gap the D+iD gap with a smaller amount of impurities
displays a reduced coherence peak reflecting full opening of the
gap around the FS. Putting together our theoretical results, we
conclude that the D+iD symmetry gap with unitary impurities
provides the most consistent agreement with the currently
available nuclear spin-lattice relaxation data.  We also studied
the effects of Born limit impurity on both symmetry gaps. We find
that
 the response to Born limit impurities  is qualitatively
different for D+iD and D-wave symmetry gaps, and both cases cannot
fit the experimental data.

\section{Formalism}

The effect of the impurity scattering is included with T-matrix
approximation\cite{Hirschfeld}. As we explained above, for
sign-changing OP and assuming particle-hole symmetric band, we
just need to calculate $\Sigma_0 =\Gamma T_0$, where
$\Gamma=n_i/\pi N_{0}$, $N_0$ the normal DOS at the Fermi energy,
$n_i$ the impurity concentration. $T_0$ is defined in Matsubara
frequency as
\begin{equation}
 T_0 (\omega_n) =\frac{g_0 (\omega_n)}{[c^2-g_0 ^2
(\omega_n)]},
\end{equation}
and
\begin{equation}
 g_0 (\omega_n) = \frac{1}{\pi N_0}  \sum_k
\frac{\tilde{i \omega}_n}{\tilde{\omega}_n^2 + \epsilon_k^2
+\Delta^2(k)},
\end{equation}
where $\tilde{\omega}_n=\omega_n+\Sigma_0$ and the scattering
strength parameter $c$ is related to the s-wave phase shift
$\delta$ as $c=\cot(\delta)$.
With this $T_0$ the following gap equation is solved
self-consistently.
\begin{eqnarray}
\Delta(k) &=& - N_0 \sum_{k'} V(k,k')  \nonumber
\\ & & \times  T \sum_{\omega_n} \int^{\omega_D}_{-\omega_D} d \epsilon \frac{
\Delta(k^{'})}{\tilde{\omega}_n^2 + \epsilon^2 +\Delta^2(k^{'})}.
\end{eqnarray}
For D+iD gap equation, we assume the pairing potential $V(k,k')$
to be a constant because thermodynamic properties of D+iD SC are
determined by the absolute magnitude of the gap $|D+iD|$ which is
the same as the isotropic s-wave gap. The property of the D+iD gap
in the above gap equation manifests itself only as the absence of
$\Sigma_2$ renormalization. For comparison, we also calculate the
D-wave case with a proper pairing potential \cite{Bang}, which
also has no $\Sigma_2$ renormalization due to the sign-changing
OP. The only technical difference of the impurity effects on D+iD-
and D-wave symmetry gaps is the normal self-energy correction due
to impurities in the SC states with a full gap and lines of nodes
in the gap, respectively.

With the gap function $\Delta(k)$ and $T_0 (\omega)$ obtained from
Eq.(1) ($T_0 (\omega)$ is analytically continued from $T_0
(\omega_n)$ by Pad\'e approximant method), we calculate the
$1/T_1$ nuclear spin-lattice relaxation rate
\cite{Hirschfeld,Choi}
\begin{eqnarray}
\frac{1}{T_1 T} &\sim&  \int_0 ^{\infty} \frac{\partial f_{F}
(\omega)}{\partial \omega}   [ (\langle Re
\frac{\tilde{\omega}}{\sqrt{\tilde{\omega}^2-\Delta^2(k)}}
\rangle_{k})^2 \nonumber \\ && + ( \langle Re
\frac{\Delta(k)}{\sqrt{\tilde{\omega}^2-\Delta^2(k)}}
\rangle_{k})^2 ],
\end{eqnarray}

\noindent where $\tilde{\omega}=\omega+\Sigma_0(\omega)$ and
$\langle...\rangle$ means the average over the FS. The first term
in the bracket of Eq. (4) is  $N^2 (\omega)$. The second term
vanishes in our calculations because of the symmetry of the OP. To
calculate $1/T_1 T$ using Eq.(4), we need the full temperature
dependence of the gap function $\Delta(k)$. The gap equation Eq.
(3) is basically the BCS gap equation, therefore it gives  the BCS
temperature dependence for $\Delta(k)$ and $\Delta_0/
T_{sc}=1.764$ and $2.14$ for the D+iD- and D-wave gap solutions,
respectively. We use the phenomenological formula,
$\Delta(k,T)=\Delta(k,T=0)~ \Xi(T)$;  $\Xi(T)=\tanh (\beta
\sqrt{T_{sc}/T-1})$, with parameters $\beta$ and $\Delta_0/
T_{sc}$. The temperature dependence of $\Sigma_0(\omega,T)$
($=\Gamma~ T_0 (\omega,T)$) is similarly extrapolated:
$T_0(\omega,T)=T_0(\omega,0)~ \Xi(T) + T_{normal}(1-\Xi(T))$,
where $T_{normal}=\Gamma/(c^2+1)$ is the normal state $T_0$. Then
we only need to calculate $\Delta(k)$ and $T_0$ at zero
temperature.

\section{Results and Discussions}
In our numerical calculation, the BCS value $\beta=1.74$  and the
ratio $2 \Delta_0/ T_{sc}= 3.5$ for D+iD gap and $2 \Delta_0/
T_{sc}= 5$ for  D-wave gap are used, respectively. Note that the
final results are insensitive to the precise value of $\beta$.
Fig.1(a) shows the normalized nuclear spin-lattice relaxation rate
1/T$_1$ for D+iD SC gap with varying impurity concentration of
unitary impurities (c=0). The normalization is chosen so that the
value of 1/T$_1$ at T$_c$ is 100 in arbitrary units. The
experimental data (black circles) are also normalized in the same
fashion. The inset shows the corresponding DOS. The key result we
find is that although D+iD gap develops a full gap as in an
isotropic s-wave gap, the unitary impurity scattering produces
resonance states inside the gap (at $\omega=0$ in our case of c=0
and a particle-hole symmetric band). A finite amount of unitary
impurities, therefore, quickly produces a residual DOS around
$\omega=0$ inside the gap. As a result, the nuclear spin-lattice
relaxation rate 1/T$_1$ develops a growing  T-linear region at low
temperatures with increasing impurity concentration. The
high-temperature 1/T$_1$ below T$_c$ initially follows the pure
D+iD curve (black open squares) with a small amount of impurities,
but with increasing impurity concentration it follows a nongeneric
power law dependence \cite{powerlaw}. With the impurity
concentration of $\Gamma/ \Delta_0=0.32$ ($\Delta_0$ is the
absolute magnitude of the pure D+iD gap at zero temperature), the
residual DOS at $\omega=0$ N(0) is about 0.63 N$_0$ (N$_0$ is the
normal state DOS at Fermi level). Compared  with the experimental
data of $^{59}$Co 1/T$_1$ from T. Fujimoto et al. (Ref.[8]) it
provides a reasonably good fit:  the low temperature T-linear
behavior and its magnitude, the high temperature power law
(experimental fit was T$^{2.2}$), and the absence of the coherence
peak. Therefore, the experimental observation of  the substantial
region of T-linear behavior \cite{Fujimoto} in $^{59}$Co 1/T$_1$
of Na$_{0.35}$CoO$_{2} \cdot y$H$_2$ O  doesn't rule out the
possibility of D+iD gap in this compound.

Furthermore, our calculations for D+iD gap provide a resolution
for the conflicting observation of the coherence peak by other
groups \cite{Kobayashi,Waki}. First, without impurities (black
open squares) the height of the coherence peak is about 1.46 times
of the value of  1/T$_1$ at T$_c$, which is a little smaller than
the value for isotropic s-wave case. This is because the second
term in Eq.(4) vanishes due to the symmetry of the D+iD OP. With
impurities the coherence peak is quickly suppressed and  there is
a small peak visible with a smaller amount of impurities (for
example, see $\Gamma/ \Delta_0=0.04$ (open purple squares); note
the logarithmic scale). We notice that there is a systematic trend
of the height of the coherence peak with T$_c$ of the samples used
for $^{59}$Co 1/T$_1$ measurement: no coherence peak with
T$_c$=3.9 K \cite{Fujimoto}; coherence peak with height 1.1 times
of the value of   1/T$_1$ at T$_c$ with T$_c$=4.5 K
\cite{Kobayashi}; coherence peak  height 1.4 times of the value of
1/T$_1$ at T$_c$ with T$_c$=5.0 K \cite{Waki}. Therefore, we can
understand the seemingly inconsistent experimental reports on
$^{59}$Co 1/T$_1$ by different groups with D+iD gap and as a
result of the different samples with different scattering
rates\cite{comment2}. We also show the D-wave results  with
unitary impurities (c=0) in Fig.1(b). The overall shape of DOS for
D-wave gap with and without impurities is qualitatively different
from the case of D+iD gap. Nevertheless, the accumulation of the
residual DOS N(0) with impurities are very similar in magnitude,
which is the main property determining the low temperature 1/T$_1$
behavior. As a result, we can also fit the data of Fujimoto et al.
with a D-wave gap with a similar amount of impurities (the
necessary $\Gamma/ \Delta_0$ value should be  between 0.32 and
0.16, see Fig.1(b)); the residual DOS N(0) with $\Gamma/
\Delta_0$=0.32 is about 0.73 N$_0$. However, the D-wave results
cannot explain the observation of the coherence peak seen by some
experimental groups \cite{Kobayashi,Waki}. For completeness, we
show the same calculations in the  Born limit  (c=1) in Fig.2(a)
and Fig.2(b). Both D+iD and D-wave gaps with Born limit impurities
doesn't produce a close fit to the experimental data unless
$\Gamma/\Delta_0 \rightarrow 0.8$. Even with $\Gamma/\Delta_0 =
0.8$, the detailed line shape of the calculated 1/T$_1$ doesn't
quite fit the experimental data for both gaps. Moreover, the
reduction of T$_c$ and the gap magnitude $\Delta_0$ with
$\Gamma/\Delta_0 = 0.8$ is about 50 $\%$ or more for both gap
cases, which is not consistent with experiments.

\section{Conclusion}

In summary, we have calculated the nuclear spin-lattice relaxation
rate 1/T$_1$ for both D+iD- and D-wave superconducting states with
impurities. We found that  unitary impurities produce the residual
density of states inside the gap due to the resonant scattering
for both cases. Interestingly, we found that the rate of
accumulation of the residual density of states with unitary
impurities is similar for both SC gaps.
As a result, both D+iD and D-wave gaps produce the T-linear
behavior in 1/T$_1$ at low temperatures and can fit the
experimental data of Fujimoto et al. \cite{Fujimoto} with the
unitary impurity concentration of $\Gamma/ \Delta_0 \sim 0.3$.
However, considering observations of the coherence peak in
$^{59}$Co 1/T$_1$ by other groups \cite{Kobayashi,Waki} and a
systematic trend of the height of the coherence peak with T$_c$ of
the samples, we conclude that the D+iD symmetry gap is the most
consistent  with the currently available nuclear spin-lattice
relaxation data of Na$_{0.35}$CoO$_{2} \cdot y$H$_2$ O.

\section{Acknowledgements}

We thank  Prof. G.-q. Zheng for  providing the experimental data
and discussions, in particular, about the details of experimental
data and interpretation and Dr. J.X. Zhu and Prof. B. S. Shastry
for discussions. This work was supported by US DoE. Y.B. was
partially supported by the Korean Science and Engineering
Foundation (KOSEF) through the Center for Strongly Correlated
Materials Research (CSCMR) (2002) and through the Grant No.
1999-2-114-005-5.

\vspace{2cm}
\begin{figure}
\epsfig{figure=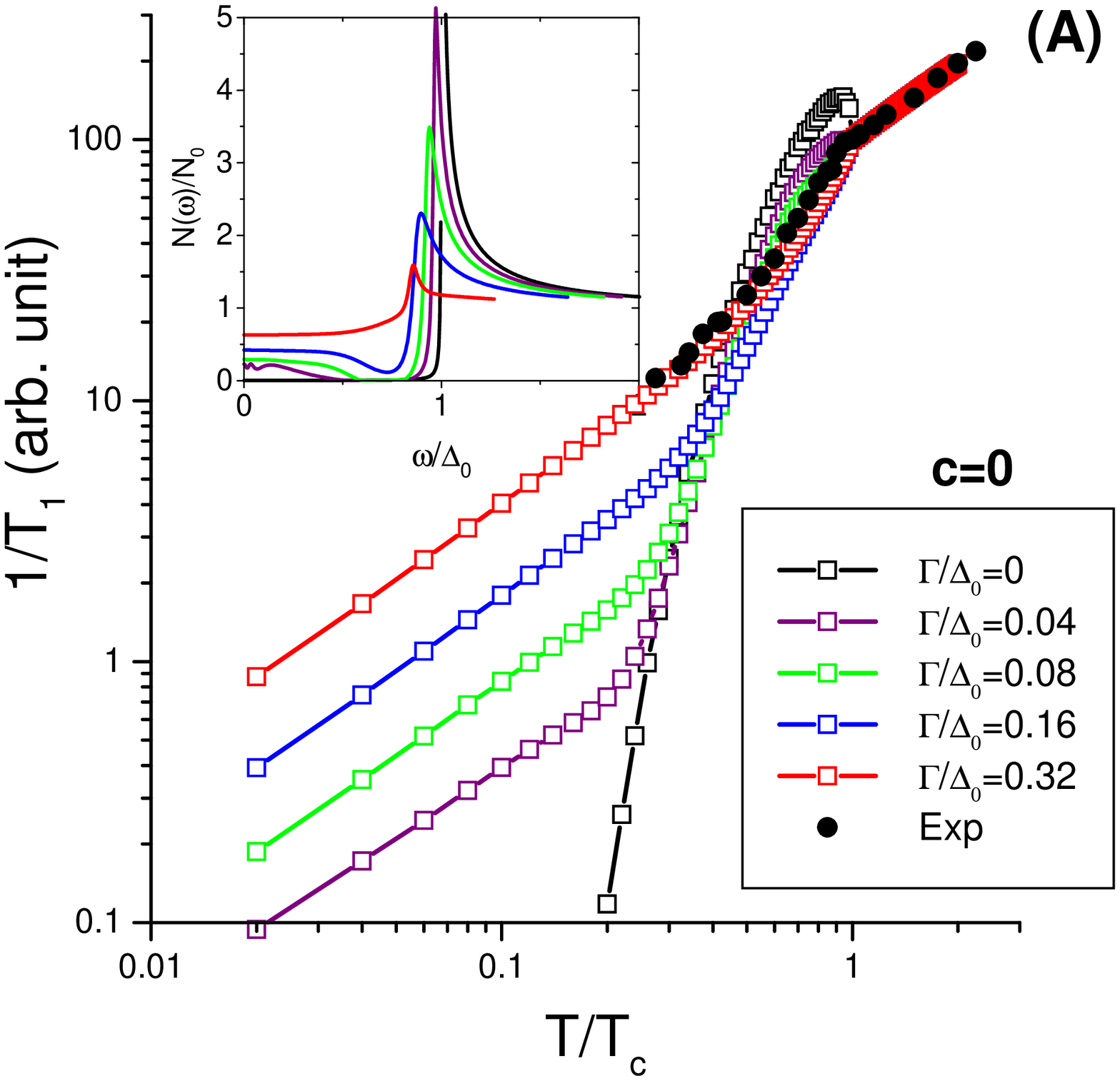,width=1.0\linewidth}
\epsfig{figure=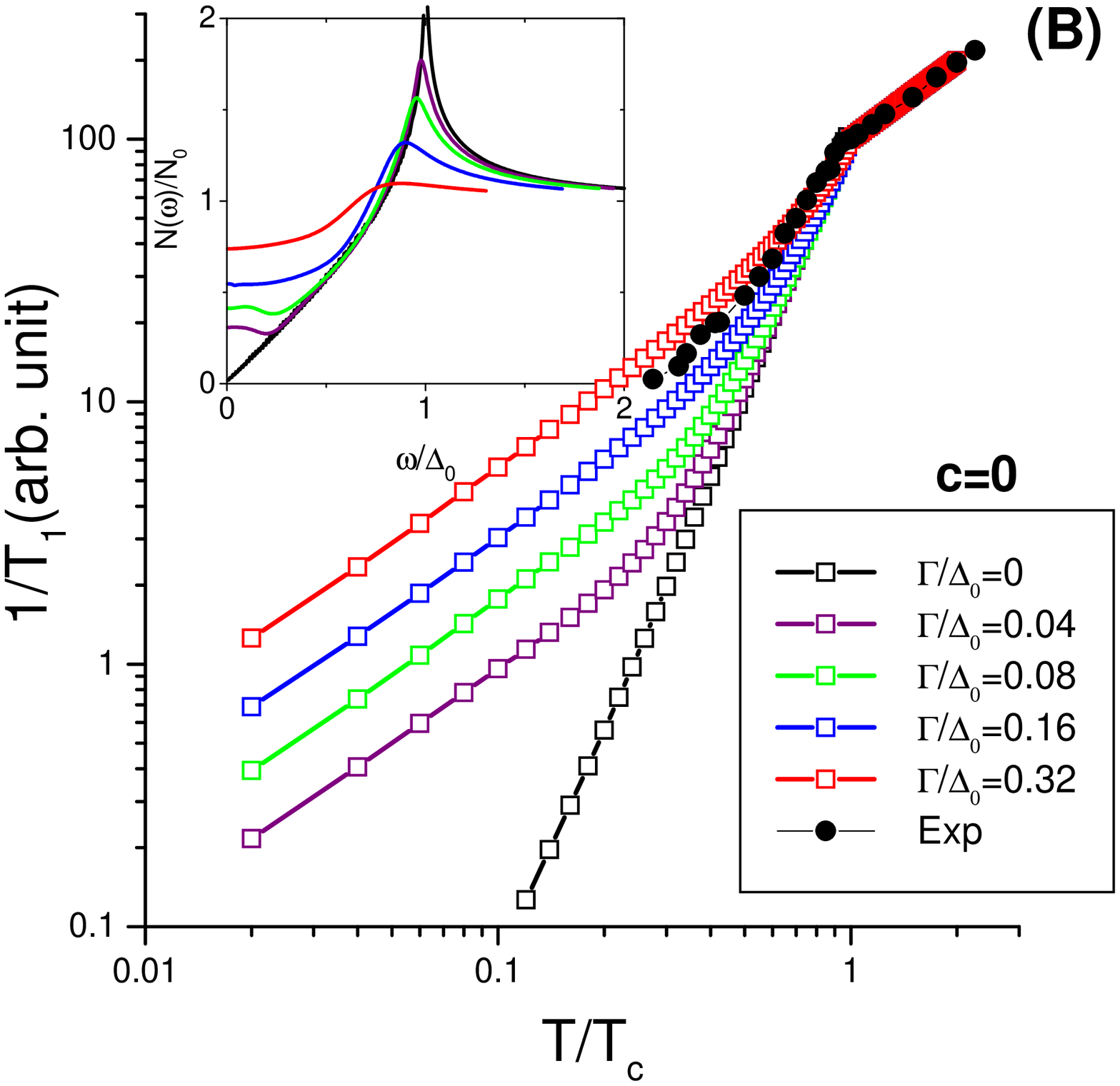,width=1.0\linewidth}
 \vspace{-1cm}
\caption{(A) The normalized 1/T$_1$ of D+iD gap with unitary
impurities (c=0).  Experimental data (black circles) are also
normalized (T. Fujimoto et al. (Ref[8])). Inset is the
corresponding DOS. (B) The same plots as in (A) for  a  D-wave gap
with unitary impurities (c=0). \label{fig1}}
\end{figure}
\vspace{2cm}
\begin{figure}
\epsfig{figure=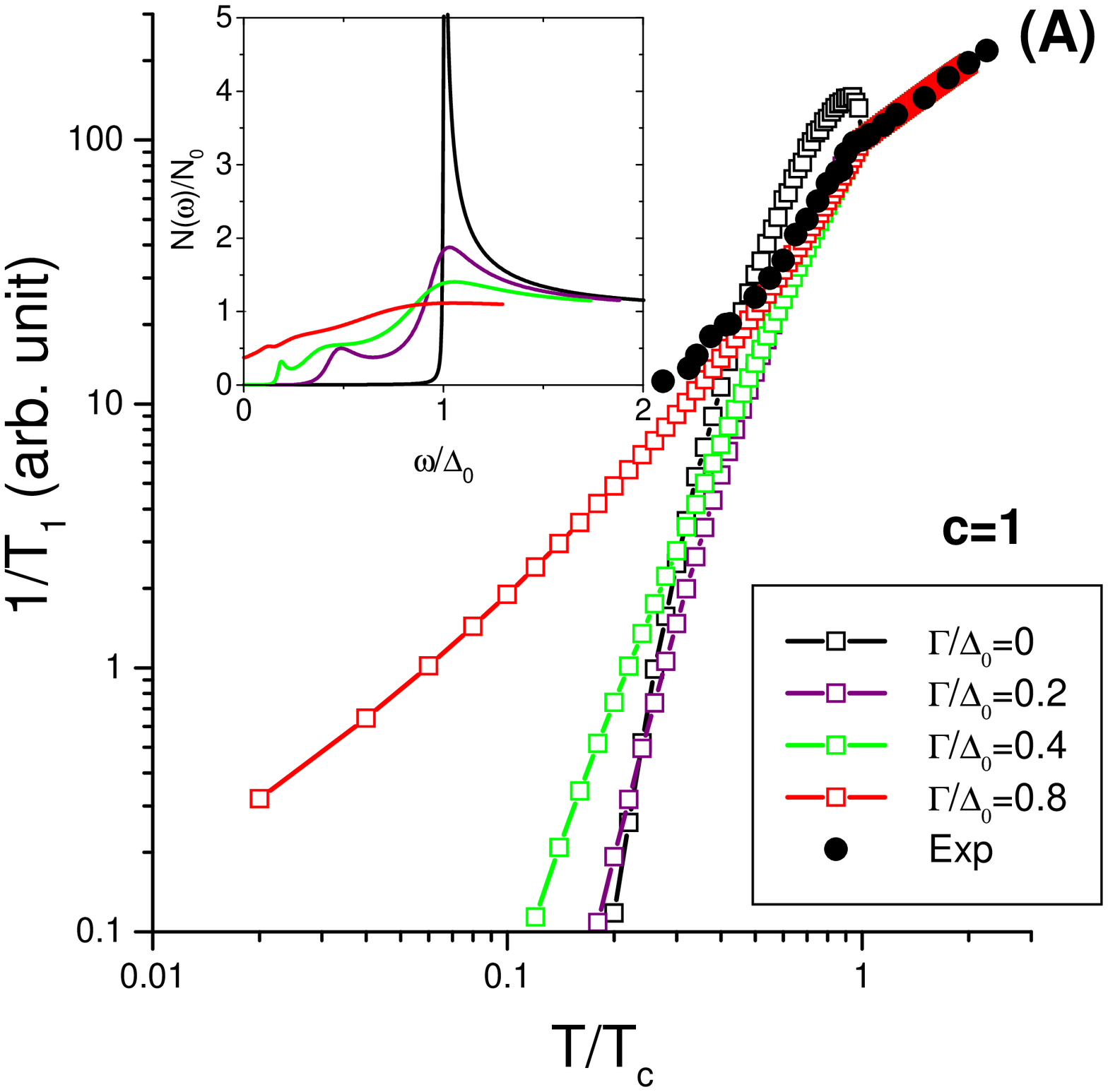,width=1.0\linewidth}
\epsfig{figure=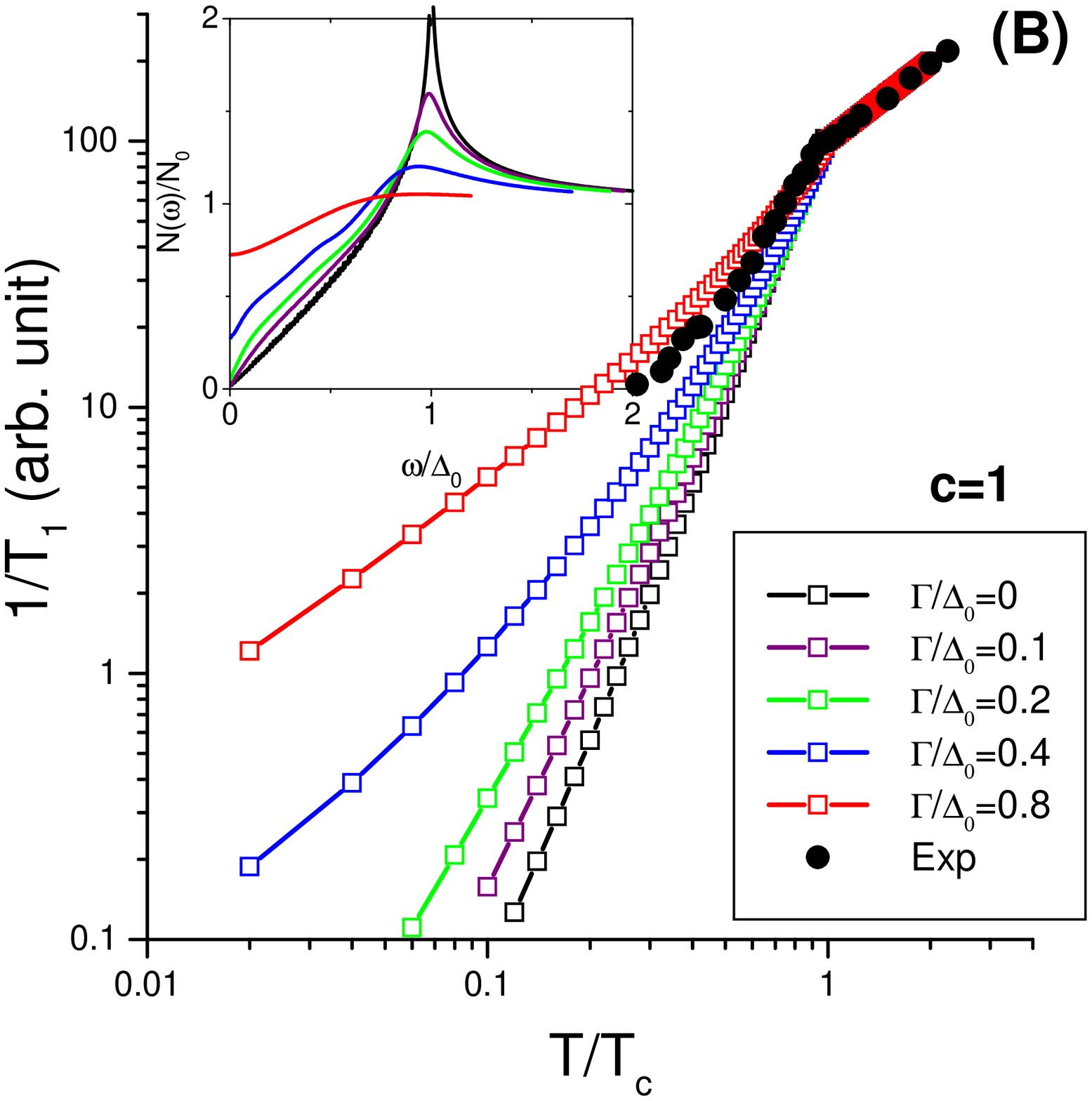,width=1.0\linewidth}
 \vspace{-1cm}
\caption{(A) The normalized 1/T$_1$ of D+iD gap with Born limit
impurities (c=1).  Experimental data (black circles) are also
normalized (T. Fujimoto et al. (Ref[8])). Inset is the
corresponding DOS. (B) The same plots as in (A) for  a  D-wave gap
with Born limit impurities (c=1). \label{fig2}}
\end{figure}

\end{multicols}

\begin{references}
\bibitem[*]{byline} Permanent address: Department of Physics, Chonnam National
University, Kwangju 500-757, Korea.
\bibitem{discovery}
K. Takada, H. Sakurai, E. Takayama-Muromachi, F. Izumi, R.A.
Dilanian and T. Sasaki, Nature {\bf 422}, 53 (2003).
\bibitem{spin_liquid}
G. Baskaran, cond-mat/0303649; Brijesh Kumar and B. Sriram
Shastry, cond-mat/0304210;  Qiang-Hua Wang, Dung-Hai Lee and
Patrick A. Lee, cond-mat/0304377.

\bibitem{Anderson}
P.W. Anderson, Science {\bf 235}, 1196 (1987); G. Baskaran, Z.
Zou, and  P.W. Anderson, Sol. St. Commn, {\bf 63}, 973 (1987).

\bibitem{D+iD}
Masao Ogata, cond-mat/0304405; Jian-Xin Zhu and A. V. Balatsky,
cond-mat/0306253.
\bibitem{Kobayashi}
Y. Kobayashi, M. Yokoi and M. Sato,  cond-mat/0305649.
\bibitem{Waki}
T. Waki, C. Michioka, M. Kato, K. Yoshimura, K. Takada, H.
Sakurai, E. Takayama-Muromachi and T. Sasalki, cond-mat/0306036.
\bibitem{Tinkham}
M. Tinkham, p 83, {\it Introduction to Superconductivity},
McGraw-Hill, Inc. (1996).
\bibitem{Fujimoto}
T. Fujimoto, G.-q. Zheng, Y. Kitaoka, R.L. Meng, J. Cmaidalka and
C.W. Chu, cond-mat/0307127.

\bibitem{Hirschfeld}
P. J. Hirschfeld, P. Woelfle, and D. Einzel, Phys. Rev. B {\bf
37}, 83 (1988).

\bibitem{Norman}
L. J. Buchholtz and G. Zwicknagl, Phys. Rev. B {\bf 23}, 5788
(1981); L. S. Borkowski and P. J. Hirschfeld Phys. Rev. B {\bf
49}, 15404 (1994); R. Fehrenbacher and M. R. Norman, Phys. Rev. B
{\bf 50}, 3495 (1994).

\bibitem{Bang}
Yunkyu Bang, I. Martin, and A.V. Balatsky, Phys. Rev. B {\bf 66},
224501 (2002).


\bibitem{Choi}
Han-Yong Choi, Phys. Rev. Lett. {\bf 81}, 441 (1998).

\bibitem{powerlaw}
In fact, the temperature dependence of $1/T_1$ at high
temperatures near T$_c$ is not a generic property  for any
symmetry gaps; it is determined by the temperature slope of
$\Delta(T)$ near T$_c$, elastic and inelastic scattering rates,
and the anisotropy of the gap, etc. Often quoted T$^3$ behavior of
$1/T_1$ for a pure D-wave gap is a generic property at low
temperatures reflecting the lines of nodes, but the same T$^3$
behavior observed at high temperatures is a result from combined
effects mentioned above. Nevertheless, empirically once the
coherence peak is suppressed the high temperature power law of
$1/T_1 \sim T^{\alpha}$ can be fit with $\alpha$ between 2 and 3
for the dirty D+iD and D cases.

\bibitem{comment2}
This conclusion needs further experimental verification since the
measurements of Ref.[5] and Ref[6] have some ambiguity in extracting
the T$_1$
relaxation time.


\end{references}
\end{document}